**Manuscript title:** **MBFormer: Microbubble Transformer for 3D Time-Series Data Processing to Improve Bound Bubble Detection in Nondestructive Ultrasound Molecular Imaging**

**Authors' names:** Jihye Baek[1], Jeong Hoon Lee[1], Hoda Hashemi[1], Arutselvan Natarajan[1], Farbod Tabesh[1], Ramasamy Paulmurugan[1], and Jeremy J. Dahl[1]

**Institutional affiliations:** [1]Department of Radiology, Stanford University School of Medicine, Stanford, CA, USA

**Corresponding author:** Jeremy J. Dahl

Stanford University

3155 Porter Dr., Palo Alto, CA 94304

Phone: 650-721-6921

Fax: 650-724-3022

Email: jjdahl@stanford.edu

**Abstract**

Development of nondestructive ultrasound molecular imaging (UMI) is essential for early cancer detection through real-time, free-hand screening using clinical ultrasound systems. Current techniques face challenges in accurately detecting targeted microbubbles (MBs) bound to specific biomarkers, primarily due to false-positive detections of unbound free-floating MBs. We propose a transformer model designed for time-series data processing, aimed at improving the differentiation of bound MBs in UMI.

We propose a hierarchical transformer, termed "**MBFormer**" (microbubble transformer), featuring a positional-embedding-free encoder and a lightweight decoder. This model leverages attention mechanisms within 3D spatio-temporal data to effectively capture stationary signals from bound MBs while suppressing non-stationary signals from unbound MBs. Given that MBs manifest as relatively small textures compared with conventional segmentation targets in medical imaging, such as organs and tumors, we optimized the model with two hierarchical layers, each with an attention block, to process spatio-temporal (ultrasound video) data. The network outputs the molecular signal amplitude to visualize fine MB textures.

Performance evaluation was conducted using an *in vivo* animal model of breast cancer, with comparisons made against a prior convolutional neural network (CNN)-based UMI method and SegFormer3D, a representative 3D transformer backbone. The results indicate that MBFormer (AUC = 0.943) outperformed both CNN (AUC = 0.897) and SegFormer3D (AUC = 0.766) in detecting bound MBs. The CNN exhibited residual molecular signal from free-floating MBs in the cardiac chambers, whereas SegFormer3D baseline failed to detect fine MB textures effectively. Overall, MBFormer demonstrated enhanced detection of bound MBs while suppressing free-floating MB signals. Furthermore, MBFormer achieved a frame rate of 16.7 to 18.1 frames per second, demonstrating its potential for real-time application. We anticipate that this improved transformer-based UMI model can facilitate real-time, free-hand nondestructive UMI in clinical systems.

## 1. Introduction

Ultrasound molecular imaging (UMI) is a promising technique designed to enhance the detection of disease through the use of contrast agents targeted to specific biomarkers of diseases[1, 2]. For example, microbubbles (MBs) designed to target biomarkers of cancer—such as the B7 homolog 3 protein (B7-H3) protein that is associated with breast cancer [3] — can significantly improve the sensitivity and specificity compared to conventional B-mode ultrasound [4]. In UMI, the targeted MBs bind to the biomarkers expressed on the endothelial cells located within the vasculature at the site of the disease and are held stationary at that location. Signals from the targeted MBs bound to biomarkers are then detected by their non-linear reflections, which can be differentiated from the linear ultrasound reflection from tissues using contrast-enhanced ultrasound (CEUS) pulse sequences (e.g. pulse inversion, pulse amplitude modulation, etc. [5-7]), thereby providing a connection between the molecular expression of a disease and the signal observed on ultrasound.

An increasing number of approaches have been developed to enhance targeted MB detection beyond the capabilities of CEUS, which often suffer from strong echo leakage from tissues and are unable to differentiate between MB that are bound to biomarker from those that are not bound. The state-of-the-art UMI is differential targeted enhancement (DTE) [8-10], which captures images before and after a bursting pulse is used to destroy the MBs in the field of view. This technique averages pre- and post-burst images separately and then subtracts the post-burst image from the pre-burst image to highlight the differences between the two. Under ideal conditions, the averaging produces a mean signal from moving microbubbles that is eliminated by the subtraction, while the bound MB signal is highlighted because it is only present in the pre-burst image. At the same time, effective suppression of tissue leakage signal is achieved because it is also present in both pre- and post-burst images. While effective, DTE does not allow for real-time imaging because of the need for the transducer to remain stationary over the target during the pre-burst, burst, and post-burst activity. In addition, the destruction of the targeted MBs leaves "holes" in subsequent images thereby preventing overlapping image captures or multiple images of the same target. Destruction of MBs also induces inertial cavitation and causes micro-damage to tissue, which is an undesirable bioeffect for a diagnostic imaging mode [11]. Consequently, we and others have developed non-destructive UMI approaches leveraging signal processing and neural networks as viable solutions to facilitate real-time and repeated ultrasound molecular imaging [12-16].

We have recently established convolutional neural network (CNN)-based nondestructive UMI approaches [12-14, 17], showing the potential of CNNs to facilitate nondestructive UMI, demonstrating performance comparable to that of the preclinical destructive DTE [13, 14, 17] and enabling its translation to clinical ultrasound systems [13, 17]. This enhanced CNN-based UMI has been validated using an in vivo animal model of breast cancer, revealing its superior performance in MB detection compared to DTE, especially in simulated free-hand scanning.

Despite these advancements in nondestructive MB detection, which addresses some of the limitations of the state-of-the-art DTE, the CNN-based framework does not fully eliminate free-floating MBs, due to its frame-by-frame processing design, thereby producing non-specific molecular signal and resulting in false-positive detection of molecular biomarkers. Consequently, there is a clear necessity of further advance the nondestructive UMI with the objective of effectively suppressing the signals generated by free-floating MBs while simultaneously improving the detection of bound MBs.

In this study, we have developed a novel transformer-based UMI model, named microbubble transformer (MBFormer), which takes in spatio-temporal 3D data (ultrasound videos) and outputs spatio-temporal 3D molecular images (UMI videos). By processing consecutive ultrasound frames, MBFormer leverages the temporal behavior of MBs: bound MBs remain stationary within the lesion, whereas free-floating (unbound) MBs move between frames. The spatio-temporal processing of MBFormer enables accurately detection of bound MBs, producing the molecular signal amplitude, while suppressing free-floating MB signals. MBFormer builds upon SegFormer3D [18] backbone to identify small and fine textures from molecular signals in UMI. Unlike conventional medical imaging applications that primarily focus on tumor or organ detection, MB detection necessitates a more sophisticated approach to accurately identify smaller and finer structures, thus encoding high-resolution features. We evaluated the performance of MBFormer by comparing it against a prior CNN-based nondestructive approach and the SegFormer3D baseline.

## 2. Related Work

### A. Neural Network for Nondestructive Real-time Ultrasound Molecular Imaging

The inference phase of our prior CNN-based UMI operates in a nondestructive and real-time manner by processing single-frame inputs of B-mode and CEUS images. However, this frame-by-frame processing of the CNN-based framework encounters difficulties in suppressing signals from free-floating MBs, which generate time-varying echoes. Enhancing the suppression of free-floating MBs is particularly achievable by processing time-series data, which carries the temporal variations in signals produced by the movement of free-floating MBs. Incorporating a temporal sequence of frames rather than individual frames has previously been shown to better discriminate between stationary (from bound MBs) and non-stationary signals (from free-floating

MBs) [12]. We anticipate that enhancing the network design by incorporating time-series data will further improve the suppression of free floating MBs and detection of bound MBs, ultimately facilitating more accurate and reliable UMI.

**B. Medical Image Segmentation and Transformers**

Semantic segmentation, enabled through pixel-wise classification via deep learning, has shown promise in advancing medical imaging techniques, facilitating the identification of anatomical structures and pathological findings, such as tumors, organs, and tissues. The advanced segmentation assists clinicians in the more effective interpretation of medical images, thereby enhancing diagnostic capabilities.

Advanced networks such as U-Net and transformers have been developed, facilitating pixel-wise segmentation for tumor detection or organ delineation. U-Net, developed by Ronneberger et al. [19] for medical image segmentation, introduced an encoder-decoder structure with hierarchical layers and skip connections. The U-Net analyzes features at multiple resolutions while retaining spatial information at each layer, thus enabling high-resolution segmentation. The U-Net based architectures quickly became the standard for medical image segmentation tasks, contributing to advancements in various applications, including tumor segmentation in ultrasound imaging [20, 21]. More recently, transformers, which utilizes attention mechanisms, have revolutionized the performance of natural language processing compared to convolutional-based networks [22]. By adapting attention-based mechanism to computer vision, the vision transformer [23] and video vision transformer [24] have been developed for image and video classification tasks, respectively. Furthermore, efforts to enhance transformer performance have resulted in the adaptation of hierarchical structures inspired by U-Net, leading to developments such as TransUnet [25], Swin Transformer [26], and Swin-unet [27].

These recent models incorporating both attention mechanisms and hierarchical structures have improved segmentation tasks. First, the attention mechanism allows for the simultaneous processing of multiple patches from input images or videos, which captures broader dependencies and global contextual relationships compared to traditional CNNs that extract local features within convolutional windows. Second, the hierarchical structure, inspired by the encoder-decoder design of U-Net, enhances model efficacy by extracting multi-scale features across resolution levels, compared to the single-scale representations of originally proposed transformers [22, 24]. However, these sophisticated transformer models present challenges for real-time applications due to their substantial memory requirements and computational complexity. As a result, efficient yet light-weight attention-based transformers have been designed to improve performance and efficiency, particularly for the processing of volumetric data [18, 25, 28-32]. Of these, SegFormer3D [18] exhibits the highest memory efficiency with the second-best performance as measured by Dice coefficient. SegFormer3D is a transformer model for volumetric segmentation, adapted from Seg-Former [29], which was originally proposed to enable memory-efficient transformers for image segmentation task. Specifically, SegFormer3D utilizes only 1.4 % and 10.7 % of parameters compared to the least memory efficient model of SETR PUP [32] and second-best memory efficient model of CoTr [31], respectively. This memory efficient and high-performance architecture of SegFormer3D may serve as the most suitable baseline for developing real-time ultrasound applications that require the processing of 3D volumetric data or temporal data, provided that appropriate model modifications are implemented. Attention-based segmentation models have primarily been optimized for relatively large size volumetric structures, such as organs or tumors, which contrast with our aim of detecting much smaller MBs. Therefore, a modification of the SegFormer3D design is necessary to accurately detect the fine textures of small microbubbles while taking advantage of the memory efficient and high-performance segmentation.

## 3. Methods

We propose MBFormer, a transformer-based UMI model designed to detect MBs bound to targeted biomarkers. We adapted MBFormer from Segformer3D [18] by modifying the network architecture, thereby achieving small-MB detection while enabling the fast processing required for real-time ultrasound applications. The model takes spatio-temporal (video) ultrasound data comprised of B-mode and CEUS images as input and produces a spatio-temporal output of the molecular imaging signal, as illustrated in Fig. 1.

The architecture of MBFormer comprises (1) a hierarchical transformer encoder featuring two resolution levels for applying attention mechanisms, (2) a lightweight decoder, and (3) a probability head. To produce a UMI frame represented as a molecular signal amplitude map, sliding window inference is conducted, and the SoftMax activation function is used to produce molecular signal amplitude maps.

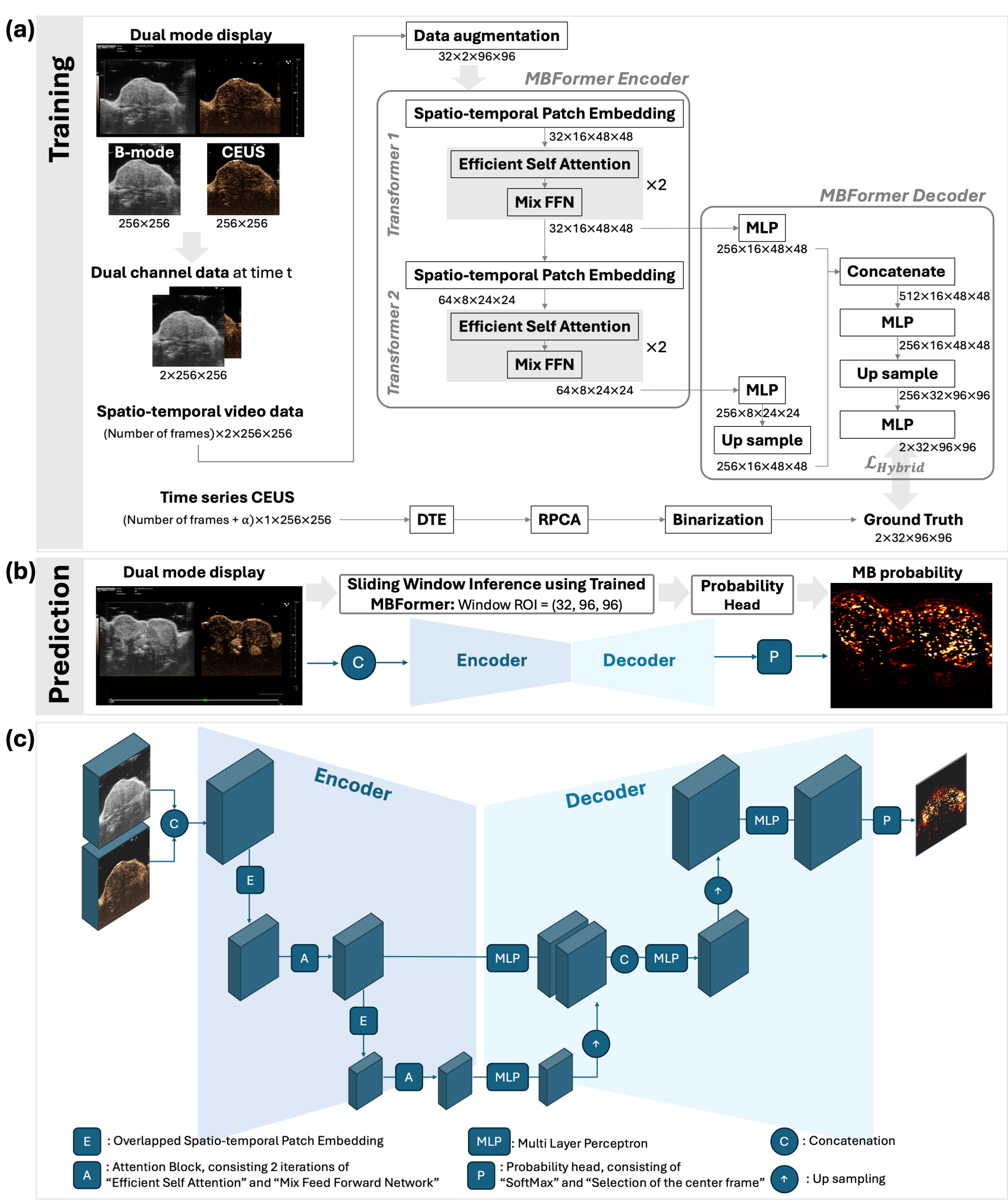

**Figure 1.** MBFormer Overview for training and prediction. (a) MBFormer training, including preprocessing to generate network input from dual mode ultrasound videos and ground truth from pre- and post-burst video recordings, DTE, and RPCA. (b) Molecular signal amplitude prediction using the trained MBFormer, consisting of sliding window inference and probability head. (c) Diagram of the network architecture, visualizing its hierarchical structure, upsampling, non-positional embedding encoder with attention mechanism, and lightweight decoder.

**A. Transformer Encoder**

The MBFormer encoder adopts a hierarchical architecture that processes input patches of size 32× 2 ×96×96 (frames×channels×height×width); details of patch extraction and data preparation are provided in **Section 3. C**. The encoder extracts features at two resolution levels, comprising three key components.

First, spatio-temporal patch embedding, adopted from overlapped patch merging in SegFormer3D [29], extracts patches from our spatio-temporal data and applies the attention mechanism while downsampling the data, thereby producing multi-scale features that serve as inputs to the hierarchical layers. The patch embedding is performed through 3D convolutional feature extraction with a kernel size of 3 and a stride of 2, which merges information among neighboring pixels and achieves a downsampling factor of 2 simultaneously with feature extraction. We chose a downsampling factor of 2 to retain information on small and fine MB textures, whereas higher downsampling factors may lose details of MB textures. The MBFormer encoder consists of two hierarchical resolution layers, corresponding to downsampling factors of 2 and 4 (i.e., 1/2 and 1/4 of the original data size in axial and lateral directions). These are relatively lower downsampling factors compared to the factors typically used for large structure segmentation (i.g. tumors and organs) in medical imaging where downsampling factors of 4, 8, 16, or 32 are commonly used.

Second, we employ efficient self-attention [33] rather than multi-head attention used in ViT [23] to reduce computational complexity from $O(N^2)$ to $O(\frac{N^2}{R})$ where we set R = 4 and 2 for the first and second layers, respectively. Each attention block, adopted from SegFormer3D [18], comprises (1) efficient self-attention and (2) a mix-feed-forward network (mix-FFN) [29].

Third, we use the mix-FFN proposed in SegFormer [29], which enables transformer encoders to operate without the positional encoding [23]. The mix-FFN has been reported to outperform positional encoding, yielding better and more robust performance on image segmentation tasks [29].

**B. Lightweight MLP Decoder**

We implemented a lightweight multilayer perceptron (MLP) decoder that fuses the features extracted from the two hierarchical resolution layers, as shown in Fig. 1. To design the decoder, we adopted the concept of lightweight MLP decoders used in SegFormer [29] and SegFormer3D [18], which avoids a complex and computationally demanding decoding process. As described in Section 3. A., the MBFormer encoder, with two hierarchical resolution layers corresponding to downsampling factors of 2 and 4, produces multi-scale features $F_1$ and $F_2$, with tensor sizes of (32×16×48×48) and (64×8×24×24), respectively. The MBFormer decoder then decodes $F_1$ and $F_2$ to produce a molecular signal feature map. The decoding is performed as follows:

$$\hat{F}_i = \text{Linear}(C_i, C)(F_i), \text{ for all } i \in \{1, 2\}$$

$$\hat{F}_{2Up} = \text{Upsample}\left(\hat{F}_2\right), \text{ with an upsampling factor of 2}$$

$$F = \text{Linear}(2C, C)(\text{Concat}(\hat{F}_1, \hat{F}_{2Up}))$$

$$F_{Up} = \text{Upsample}(F), \text{ with an upsampling factor of 2}$$

$$M = \text{Linear}(C, N_{cls})(F_{Up})$$

where $\text{Linear}(a, b)$ denotes a fully connected (linear projection) layer that maps an input of a channels to b channels, $\text{Concat}(\cdot)$ denotes channel-wise concatenation, and $\text{Upsample}(\cdot)$ denotes upsampling. First, the multi-scale encoder features $F_1$ and $F_2$, with $C_i$ channels ($C_1 = 32$ and $C_2 = 64$), are projected to a unified decoder channel dimension $C$ ($C = 256$) through fully connected layers, yielding the linearly projected features $\hat{F}_1$ and $\hat{F}_2$ with sizes of ($C$ ×16×48×48) and ($C$ ×8×24×24), respectively. Next, $\hat{F}_2$ is upsampled by a factor of 2 to match the spatial resolution of $\hat{F}_1$, producing $\hat{F}_{2Up}$ with a size of ($C$ ×16×48×48). The two features $\hat{F}_1$ and $\hat{F}_{2Up}$ are then concatenated along the channel dimension, resulting in a tensor with size of (2$C$ ×16×48×48), which is subsequently fused through a fully connected layer to produce the feature $F$ with a size of ($C$ ×16×48×48). $F$ is further upsampled by a factor of 2, yielding $F_{Up}$ with a size of ($C$ ×32×96×96). Finally, a fully connected layer projects $F_{Up}$ to the molecular signal feature $M$ with a size of ($N_{cls}$ ×32×96×96), where $N_{cls}$ indicates the number of classes ($N_{cls}$ = 2): bound MBs (the detection target) and non-bound signals, including freely moving MBs and background.

**C. MBFormer Training and Inference**

MBFormer is designed to take patch inputs extracted from ultrasound videos through the encoder and to produce molecular feature maps through the decoder, where both the input and output tensors have spatial dimensions of (96×96). Our dual-mode display, showing B-mode and CEUS images side by side, was recorded as an MP4 video, stored as image files, and resized to (256×256) pixels using the MATLAB imresize function; further details on video conversion are provided in [13].

Specifically, the network operates on (96×96) patches during training, whereas inference reconstructs full (256×256) maps via a sliding-window approach. The B-mode and CEUS tensors, sized at (256×256), underwent the pre-patch-extraction and post-processing steps to match the spatial dimensions to (96×96), through data augmentation and a sliding-window inference ap-proach. The spatial dimension reduction from 256 to 96 was necessary because multi-layer hierarchical transformer processing of the entire volume is computationally prohibitive. Note that the dimension reduction was performed by cropping without downsampling.

The videos were pre-recorded to train the network, where the total frame numbers in each video ($N_{frames}$) range from 32 to 400. However, our inference aims at free-hand real-time application, and therefore we defined temporal processing unit to update the prediction of molecular signal amplitude map in real-time. We process the spatio-temporal data with this unit, denoted "$procFrame$", and update a single frame molecular signal map. We set $procFrame$ = 32 frames for our MBFormer architecture. Therefore, given unit spatial dimension of $(96 \times 96)$ and $procFrame$ of 32, the unit patch to feed the MBFormer encoder has the dimension of $(2, procFrame, 96, 96)$, where "2" represents the concatenated B-mode and CEUS channels.

To extract the unit patch from (2× $N_{frames}$ ×256×256) for training, we performed data augmentation. We extracted 42 patches of size $(2,\ procFrame, 96, 96)$ from each video, applying crops, flips, and rotations during extraction. The 42 folds represents the maximum augmentation rate, taking into account our dataset size, the capacity of two GPUs (Quadro RTX 8000 GPUs), and the dimensions of the network models. We randomly extracted the patches with the spatial size of $(96, 96)$ with temporal size of $procFrame$. While extracting, we applied lateral flips with a probability of 0.3 and rotation with a probability of 0.5, respectively. Because raw ultrasound signals exhibit anisotropic spatial resolution, we applied only lateral flips and further limited the rotation to $\pm$20.6°. Although the anisotropy originates in the raw radio frequency (RF) data, it is substantially reduced by the processing pipeline (including envelope detection, time-gain compensation, speckle reduction, and image post-processing) applied to generate the displayed images. This reduction of anisotropy allows small-angle rotations to be used as an augmentation strategy, increasing the diversity of the training data.

The MBFormer training was performed using our data augmentation, MBFormer encoder, and the MBFormer decoder. For inference, the trained MBFormer processes unit tensors of size ($2\times procFrame \times 256\times 256$). The MBFormer data buffer is be designed to accumulate the previous 32 frames' (as $procFrame$) B-mode and CEUS data to predict current UMI display. A sliding-window approach produces the molecular feature maps at the spatial dimension of (256×256), corresponding to the input B-mode and CEUS tensors. This approach first produces partial molecular feature maps across entire videos and then combines the partial maps with 25% overlap to generate the entire feature map of size ($procFrame \times 2\times 256\times 256$), where the channel dimension of 2 corresponds to the bound MB and non-bound MB classes. The MB prediction head then applies the SoftMax activation function to produce the molecular signal amplitude map of size (256×256). Among 32 predicted frames, the center (16th) frame is displayed for UMI. Note that generating a single UMI frame requires processing 32 frames of time-series data to analyze temporal nature of MB signals: bound MBs generate stable signals over time, whereas unbound, free-floating MBs generate time varying signals.

### D. Ground Truth and Loss Functions

The ground truth for training was generated from RPCA-filtered DTE videos. We first generated DTE videos through a sliding DTE approach, described in [13]. Each frame of the DTE videos ($DTE_i$) was obtained by:

$$DTE_i = \frac{1}{30} \sum_{k=i-15}^{i+15} CEUS_k^{sig} - \overline{CEUS_F^{ref}}$$

where $CEUS_k^{sig}$ is a signal CEUS frame, and $\overline{CEUS_F^{ref}}$ is the averaged reference CEUS frame without the presence of MBs, given by:

$$\overline{CEUS_F^{ref}} = \frac{1}{F} \sum_{k=1}^{F} CEUS_k^{ref}$$

where $CEUS_k^{ref}$ is a reference frame $k$ and F represents the number of frames used for averaging. We employed the sham, injection, and burst DTE approaches, and the selection of the signal and reference frames for each DTE type is summarized in Table 1. Further details on the DTE types can be found in [13].

**Table 1**. Selection of signal and reference frames for each DTE type

| DTE type | $CEUS_k^{sig}$ | $CEUS_k^{ref}$ [a] (reference, without MBs) |
|---|---|---|
| Sham | Later pre-injection frame | Early pre-injection frame |
| Injection | Post-injection frame | Pre-injection frame |
| Burst | Pre-burst frame | Post-burst frame |

[a] Reference CEUS frames without the presence of MBs.

The DTE videos, denoted as the set of all DTE frames $\{DTE_i\}$, were filtered using Robust Principal Component Analysis (RPCA) to differentiate between bound MBs and free-floating MBs [12], yielding:

$$DTE_i = DTE_i^{\text{bound}} + DTE_i^{\text{free}}$$

where $DTE_i^{\text{bound}}$ and $DTE_i^{\text{free}}$ represent the low-rank and sparse images, which correspond to images from bound and free-floating MBs, respectively. Each $DTE_i^{\text{bound}}$ was binarized to generate training ground truth with labels $\boldsymbol{y} \in \{0, 1\}$ where 0 and 1 represent non-bound MBs and bound

MBs, respectively. The label 0 includes free-floating MBs, tissue signals, and background. The intensities of $DTE_i^{\text{bound}}$ range from 0 to 1, whereas those of $DTE_i$ range from 0 to 255. Our prior CNN-based UMI work optimized a DTE threshold intensity of 20 to binarize $DTE_i$ to generate the ground truth for network training [13]. To determine a threshold on $DTE_i^{\text{bound}}$ corresponding to the optimized threshold of 20 on $DTE_i$, we compared the histograms of $\{DTE_i^{\text{bound}}\}$ and $\{DTE_i\}$, computed over all frames in a video, and found that the two distributions closely matched after rescaling $DTE_i^{\text{bound}}$ by a factor of 100. Accordingly, the previously optimized threshold 20 on $DTE_i$ corresponds to a threshold of 0.2 on $DTE_i^{\text{bound}}$, which we adopted for binarization to generate the network training ground truth.

We employed a hybrid loss function, denoted as $\mathcal{L}_{hybrid}$, which integrates the cross-entropy loss ($\mathcal{L}_{XEnt}$) and soft dice coefficient (SDC) loss ($\mathcal{L}_{SDC}$) functions [13]. The hybrid loss is a weighted sum of $\mathcal{L}_{XEnt}$ and $\mathcal{L}_{SDC}$ formulated by:

$$\mathcal{L}_{hybrid}(\hat{y}, y) = (1-\beta)\mathcal{L}_{XEnt}(\hat{y}, y) + \beta\mathcal{L}_{SDC}(\hat{y}, y), \quad \text{(eq 2)}$$

where $0 < \beta < 1$ is a weighting coefficient, and

$$\mathcal{L}_{XEnt}(\hat{y}, y) = -\sum_{i \in ROI}[\hat{y}_i \log \hat{y}_i + (1-\hat{y}_i)\log(1-\hat{y}_i)], \quad \text{(eq 3)}$$

$$\mathcal{L}_{SDC}(\hat{y}, y) = \frac{\sum_{i \in ROI} 2\hat{y}_i y_i + \epsilon}{\sum_{i \in ROI}(\hat{y}_i + y_i) + \epsilon} \ . \quad \text{(eq 4)}$$

Here, we used a smoothing term, $\epsilon = 10^{-10}$, and optimized $\beta$ as 0.5.

## 4. Experiments

The network models of MBFormer, the baseline SegFormer3D, and the models developed for our ablation study were implemented in PyTorch (version 2.5.1) using two NVIDIA Quadro RTX 8000 GPUs, each with 48 GB of memory. For the baseline CNN evaluation, we used the previously validated code in TensorFlow (version 2.18.0) using an Intel Core i7-12700K CPU [13].

**A. Datasets**

We previously collected *in vivo* animal data with the approval of the Institutional Administrative Panel on Laboratory Animal Care (APLAC) at Stanford University [13]. This dataset included ultrasound B-mode and CEUS videos of transgenic mice [FVB/N-TG(MMTV-PyMT$^{634Mul}$)] modeling spontaneous breast cancer, with corresponding ultrasound molecular images derived from DTE imaging. The videos in this dataset were collected using a Vevo2100 ultrasound scanner (FUJIFILM VisualSonics, Toronto, ON, Canada) equipped with an 18 MHz linear transducer (MS250S) using non-targeted (control) and targeted MBs (B7-H3, PD-L1, and VEGFR2). In total, 83 videos from 14 animals are available from this dataset through distinct MB injections and recordings of injection and burst videos. For each animal, we recorded pre-injection videos for less than 30 seconds, injected MBs, recorded post-injection videos for approximately 2 minutes, and then waited for approximately 1 minute to allow the MBs to circulate to the breast lesions. While waiting, bound MBs were attached to the biomarkers, whereas free MBs were washed out. Subsequently, we recorded pre-burst videos for 30 seconds, destroyed the MBs, and recorded post-burst videos for 30 seconds. We analyzed these videos to generate training and testing datasets and performed 5-fold cross-validation, in which the data was divided into 80% for training and 20% for testing. Further details on microbubble fabrication, dataset generation, and the data split can be found in [13].

**B. Ablation Study**

To evaluate the contribution of each main architectural component of MBFormer, we conducted an ablation study on the downsampling factors, the number of hierarchical layers, the loss functions, and the unit ensemble size (*procFrame*; i.e., the number of frames used to predict a single-frame UMI), as follows. (1) We tested the initial downsampling factors of 2 and 4, applied during the spatio-temporal patch embedding at the first hierarchical layer of the MBFormer encoder. In all subsequent layers, a fixed downsampling factor of 2 was applied at each layer transition. This ablation examined the effect of the initial downsampling factor on MB texture visualization. (2) We tested two and four hierarchical layers to investigate the trade-off between real-time capability and MB texture visualization. (3) We compared the Dice loss with a hybrid loss combining the Dice and cross-entropy losses to assess their effect on MB texture visualization. (4) We varied the unit ensemble size ($procFrame$, as defined earlier) among 8, 16, and 32 frames to investigate the trade-off between real-time capability and the suppression of free-floating MB signals.

Each component was evaluated using both an add-one-in and a leave-one-out strategy: by incorporating it into the SegFormer3D [18] baseline and by reverting it from the final MBFormer, respectively. All ablation models were compared against the CNN-based UMI [13] and SegFormer3D [18] baselines.

**C. Evaluation metrics**

Because MBFormer produces continuous molecular signal amplitudes rather than binary predictions, we evaluated bound MB detectability using metrics suited to continuous outputs, avoiding threshold-dependent binary metrics such as sensitivity and specificity. First, the Pearson correlation coefficient (PCC) was computed between the predicted molecular signal amplitude (ranging from 0 to 1) and the reference amplitude from RPCA-filtered DTE imaging. The measures were averaged within each manually contoured lesion and across frames for each video. As accurate estimation of molecular signal amplitude is the primary goal of our UMI, PCC serves as the most critical metric. Second, the area under the receiver operating characteristic curve (AUC) and the continuous Dice coefficient (CDC) [34] were computed directly from the continuous molecular signal amplitudes (without binarizing the predictions) against the binarized ground truth (Section 3. Methods, D. Ground Truth and Loss Functions).

To assess the suppression of free-floating (unbound) MBs, we computed the correlation coefficient (CC) between consecutive frames. Since bound MBs remain stationary while free-floating (unbound) MBs move between frames, higher inter-frame consistency (CC approaching 1) indicates more effective suppression of free-floating MBs and background. We additionally quantified the detected molecular signal as a function of the amount of free-floating MBs. The amounts of bound and free-floating MBs were estimated as the averaged intensities within the lesion contour from the low-rank and sparse matrices of the RPCA filter, respectively. We computed the averaged molecular signal amplitude within a lesion, $\hat{p}$, as:

$$\hat{p} = \frac{1}{M}\sum_{i \in L} p_i$$

where $p_i$ is the predicted molecular signal amplitude at pixel $i$, $L$ denotes the set of pixels within the lesion contour, and $M$ is the number of pixels within $L$.

Lastly, we measured inference time per frame to assess the maximum possible frame rate for real-time implementation of nondestructive UMI. Inference times were measured in PyTorch (version 2.5.1) using two NVIDIA Quadro RTX 8000 GPUs, each with 48 GB of memory, in parallel.

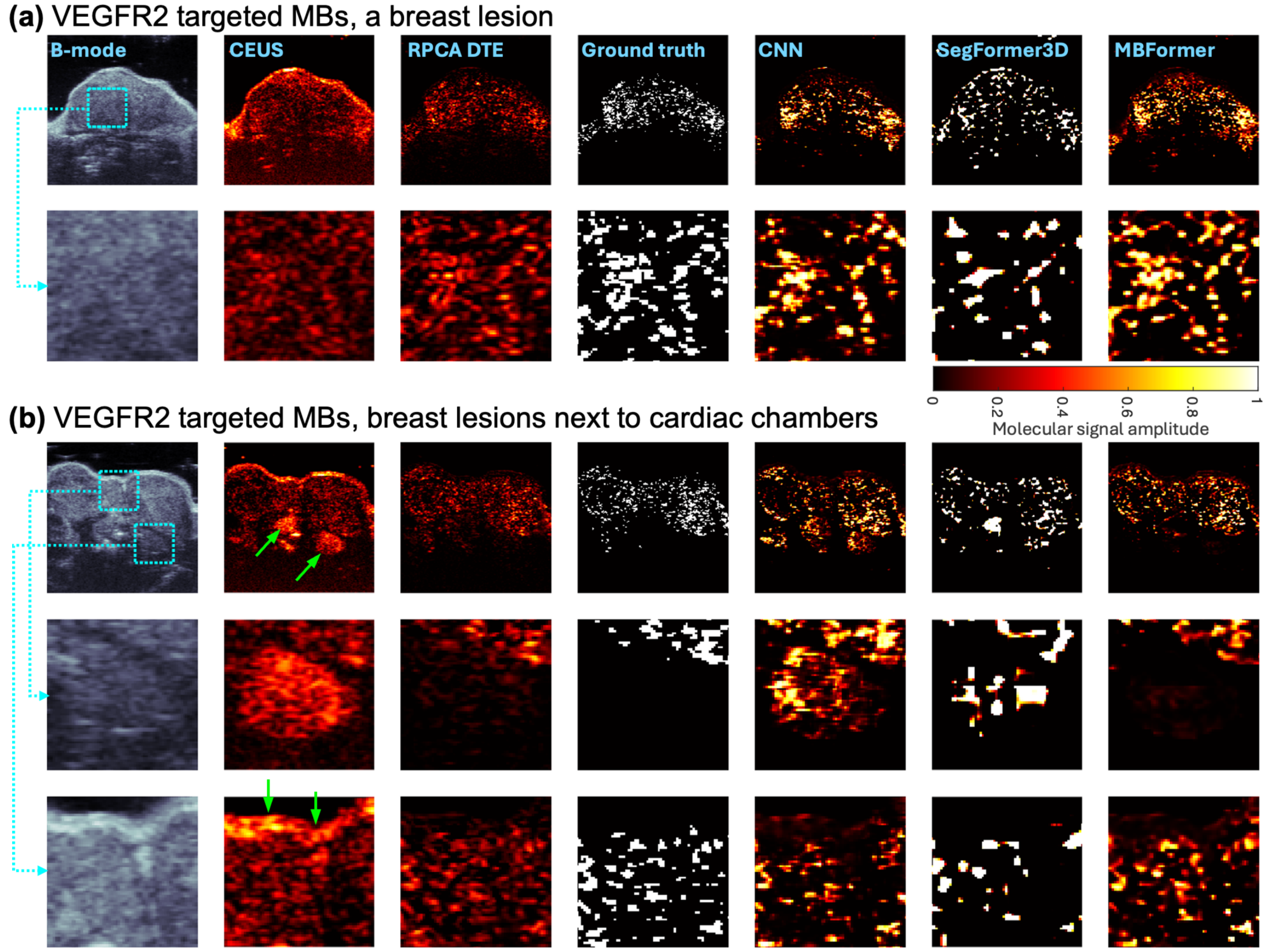

**Figure 2.** Qualitative model comparison between MBFormer, SegFormer3D, and CNN-based UMI images with insets. (a) MBFormer resulted in finer MB textures compared to SegFormer3D and CNN-based UMI as observed in the insets. (b) The insets show microbubble accumulation in the cardiac chambers (second row) and at the tissue boundary (3rd row). MBFormer suppresses the signals from free floating MBs in the cardiac chambers, whereas CNN-based UMI highlights these false molecular signals. Both nondestructive UMI (CNN and MBFormer) suppress the boundary tissues, indicated by the arrows (third row).

## 5. Results

### A. Improvement from the baselines

Fig. 2 presents the B-mode and CEUS input images, RPCA-processed DTE, ground truth generated from RPCA-processed DTE, and output molecular signal maps from CNN, SegFormer3D, and MBFormer. These images were acquired after injecting VEGFR2-targeted MBs but before destroying the MBs. Fig. 2 (a) displays a single breast lesion, while Fig. 2 (b) displays two breast lesions. Cardiac chambers are visible in Fig. 2(b) below the tumor in the CEUS image due to the high concentration of free-floating MBs in the blood pool (indicated by the arrows). In the insets in Fig. 2 (a), MBFormer showed finer MB texture that more closely resembles that seen in the DTE and ground truth images compared to CNN and SegFormer3D. Although CNN visualized MB textures better than SegFormer3D, MBFormer still resembled the DTE and ground truth more closely. In Fig. 2 (b), the cardiac chambers (second row) and tissue leakage (third row) show strong signals in the CEUS insets. While the CNN suppressed the tissue leakage signal compared to the CEUS image, the CNN still displayed false molecular signals from the cardiac chambers (indicated by arrows in the CEUS image) created by free-floating MBs. SegFormer3D also suppressed the signal from the tissue leakage, but was unable to capture fine MB texture or suppress free-floating MBs, resulting in a poor match to the DTE image. MBFormer showed comparable MB detection with the DTE and ground truth image while avoiding artifacts. Based on the qualitative results shown in Fig. 2, MBFormer demonstrated improved detection of bound MBs compared to CEUS, CNN-based UMI, and SegFormer3D.

Quantitative evaluation over all images is summarized in Figure 3. In Figure 3 (a), MBFormer outperformed both CNN and SegFormer3D on all metrics except inference time and frame rate in detecting bound MBs. Although CNN exhibited the fastest inference time, MBFormer also achieved a frame rate of 16.7 FPS, indicating its potential for real-time implementation. For CC, the transformer-based models processing spatio-temporal 3D data (ultrasound videos) achieved high inter-frame consistency, indicating stable detection of bound MBs across frames. Although CDC appeared lower than the other metrics, this reflects its inherent reduced sensitivity to small targets like MBs compared to larger structures such as organs, rather than degraded performance, as CDC is reduced for small targets compared to larger structures, such as organs. Figure 3 (b) shows receiver operating curves with AUCs. Figure 3 (c) compares molecular signal amplitude measures between RPCA-filtered DTE and the network models. MBFormer achieved the highest PCC and best matched the measured values with RPCA-filtered DTE (slope of curve fit = 1.0). Overall, MBFormer is the most competitive model in detecting molecular signal amplitudes from bound MBs.

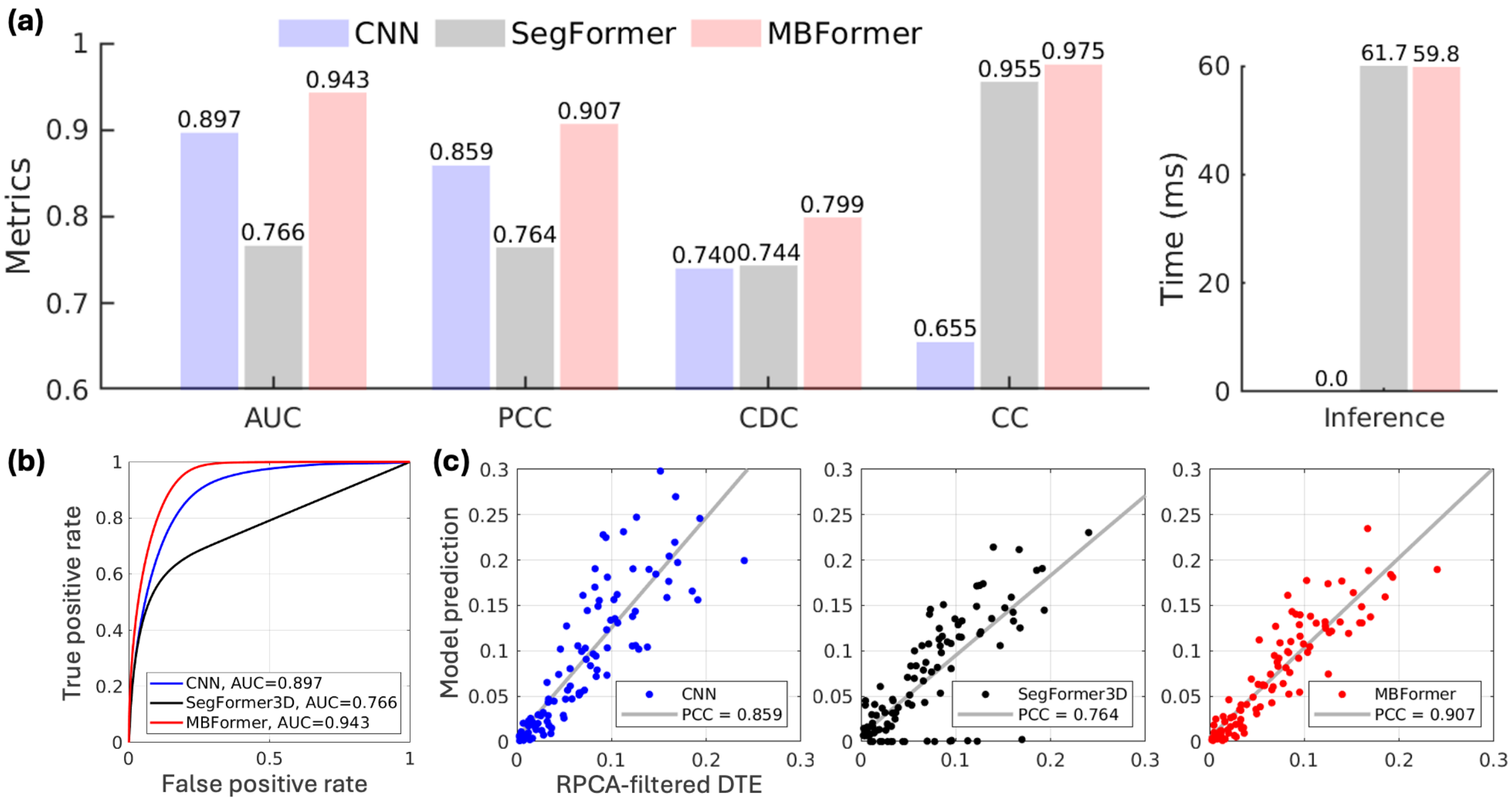

**Figure 3**. MBFormer performance comparison to CNN-based UMI and SegFormer3D. (a) Bar plots show AUC with the reference ground truth, PCC with RPCA-filtered DTE, CDC with the reference ground truth, CC between consecutive frames, and inference time per frame. (b) Receiver operating characteristic (ROC) curves averaged across the animals, with corresponding AUC values. (c) Scatter plots showing the correlation between RPCA-filtered DTE and network prediction values in each lesion. Each dot represents one video averaged across the frames, and the lines denote the linear fit to the scatter points. A slope closer to 1.0 indicates better molecular signal amplitude agreement between RPCA-filtered DTE and network prediction.

## B. Ablation study

To assess the effectiveness of the MBFormer, we evaluated the contributions of three main modifications to the SegFormer3D baseline: (1) **a downsampling factor of 2 (replacing 4)**, (2) **a** **reduction in the number of hierarchical layers** from 4 to 2; and (3) **a hybrid loss replacing the** Dice loss. We additionally examined the effect of the number of frames used in video processing (8, 16, and 32), whereas the prior CNN-based UMI processes a single frame. Starting from the baseline, each component was added individually (Add-one-in), and from the full MBFormer, each was removed individually (Leave-one-out). The results are summarized in Table 2 and Figs. 4 – 5.

**Table 2**. Ablation study

| Model Name | Model Type | Downsampling factor of 2 | 2 Hierarchical layers | Hybrid Loss | Ensemble unit | PCC with DTE[a] | AUC | CDC | CC | Inference time (ms) | Frame rate (FPS) |
|---|---|---|---|---|---|---|---|---|---|---|---|
| CNN (Our prior UMI) | UMI baseline | - | - | - | 1 | 0.86 | 0.90 | 0.74 | 0.66 | ***0.8*** | ***1250.0*** |
| SegFormer3D | Baseline | | | | 32 | 0.76 | 0.77 | 0.74 | 0.96 | 61.7 | 16.2 |
| SegFormer3D + Hybrid loss | Add-one-in | | | ✓ | 32 | 0.89 | 0.93 | 0.73 | **1.00** | 60.6 | 16.5 |
| SegFormer3D + Downsampling | | ✓ | | | 32 | 0.82 | 0.79 | **0.83** | 0.92 | 75.8 | 13.2 |
| SegFormer3D + Layers | | | ✓ | | 32 | 0.62 | 0.76 | 0.82 | 0.98 | 48.8 | 20.5 |
| MBFormer - Hybrid loss | Leave-one-out | ✓ | ✓ | | 32 | 0.73 | 0.80 | 0.82 | 0.92 | 72.4 | 13.8 |
| MBFormer - Downsampling | | | ✓ | ✓ | 32 | 0.87 | 0.93 | 0.72 | **1.00** | 60.8 | 16.5 |
| MBFormer - Layers | | ✓ | | ✓ | 32 | 0.87 | 0.94 | 0.80 | 0.98 | 76.4 | 13.1 |
| MBFormer (32 frames) | Ours | ✓ | ✓ | ✓ | 32 | **0.91** | 0.94 | 0.80 | 0.98 | 59.8 | 16.7 |
| MBFormer (16 frames) | | ✓ | ✓ | ✓ | 16 | 0.90 | **0.95** | 0.79 | 0.97 | 58.9 | 17.0 |
| MBFormer (8 frames) | | ✓ | ✓ | ✓ | 8 | 0.89 | 0.94 | 0.77 | 0.97 | 55.2 | 18.1 |

SegFormer3D serves as our baseline, adapted from the original SegFormer3D by reconfiguring its input to ultrasound 2-channel (B-mode and CEUS) video frames, following our prior CNN-based UMI. Starting from the baseline, each component is added individually (Add-one-in), and from the foll model MBFormer, each is removed individually (Leave-one-out). All models share the same 2-channel input and frame configuration for fair comparison. Bold indicates the best result per metric. Shaded cells denote high performance (PCC, AUC, CC ≥ 0.9; CDC ≥ 0.8). The lower threshold for CDC was chosen because it is inherently reduced by pixel-level misalignment in small structures, such as the microbubbles in our study. [a]PCC with DTE: pearson's correlation coefficient with RPCA-filtered DTE.

Fig. 4 displays the scatter plots summarizing the ablation models' performance. Since PCC quantifies the accuracy of predicting molecular signal amplitude, which is the most critical metric for our UMI, it serves as the x-axis in the scatter plots of Fig. 4. Fig. 4 (a) compares PCC and AUC, and Figure 4 (b) compares PCC and FPS: insets show the high-performance region for each. Fig. 4 (a) evaluates model performance in terms of capabilities of illustrating molecular signal amplitude (Fig.4(a) x-axis) and segmenting MBs (Fig.4(b) y-axis). All MBFormer configurations achieved both high PCC and high AUC (≥ 0.9), placing them in the shaded high-performance region. Fig. 4 (b) further evaluates the trade-off between model performance (with x-axis PCC) and inference time (with y-axis FPS). It shows that MBFormer maintained a competitive frame rate (16.7-18.1 FPS) while achieving the highest PCC, demonstrating a favorable balance between accuracy and real-time capability.

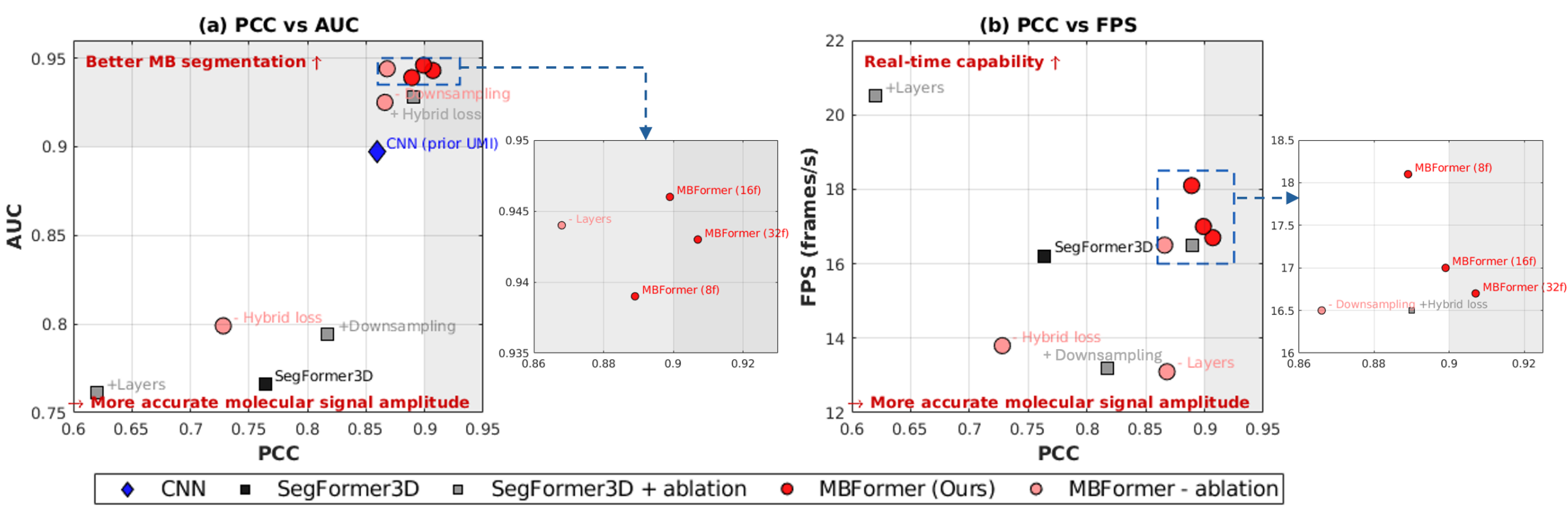

**Figure 4.** Scatter plots summarizing the ablation study. Model performance is shown with PCC on the x-axis, as it quantifies the accuracy of molecular signal amplitude (the most critical metric for our UMI) versus (a) AUC and (b) frame rate (FPS). CNN is excluded from (b) owing to its substantially higher frame rate (1250 FPS). Insets show magnified views including the performance of MBFormer configurations (dashed boxes). Shaded regions denote high performance (PCC, AUC ≥ 0.9); arrows indicate the direction of better performance along each axis. All MBFormer configurations lie within the high-performance region while maintaining a competitive frame rate.

Table 2 summarizes the performance of all ablation models. The highest performance for each metric is bold, and the shaded cells denote high performance (PCC, AUC, CC≥ 0.9; CDC ≥ 0.8). MBFormer processing 32 frames achieved the best PCC and AUC. Although the MBFormer model was not ranked first in CDC and CC, the values were comparable to the top-ranked results (CDC = 0.80 vs. 0.83; CC = 0.98 vs. 1.00). The relatively higher CDCs of the Dice-loss models do not reflect a better representation of MB texture: despite being continuous prediction values between 0 and 1, their outputs appeared near-binarized (Fig. 5), which tends to inflate CDC (CDC for ablation models with Dice loss = 0.82 – 0.83). For CC, all models processing spatio-temporal 3D data, including SegFormer3D, MBFormer, and ablation models, maintained high inter-frame consistency (≥0.92), a substantial improvement over the single-frame CNN baseline (CC = 0.66), indicating that time-series processing effectively suppresses non-stationary signals from free-floating MBs.

Fig. 5 visualizes the effects of the ablation study, showing the transition from the baseline CNN (Fig. 5 (a)) and SegFormer3D (Fig. 5 (b)) to MBFomer (Fig. 5 (c)) with the corresponding molecular signal maps. Increasing the downsampling factor from 4 to 2 (arrows labeled 'f2') consistently produced finer MB textures: (1) from (b) to (f); (2) from (h) to (i); (3) from (k) to (c). Replacing the Dice loss with the hybrid loss (arrows labeled 'Hybrid loss') enabled a more sensitive depiction of molecular signal amplitude, thereby improving the near-binarized prediction like segmentation, allowing for capturing more molecular signals within the correct regions defined by the ground truth (Fig. 5 (l)). Moreover, both the changes to the downsampling factor of 2 and the hybrid loss suppressed flow signals from the cardiac chambers. In contrast, reducing the hierarchical layers from 4 to 2 (arrows labeled '2 layers') did not noticeably affect UMI quality but increased the frame rate, such as: (i) 13.1 to (c) 16.7 FPS; (b) 16.2 to (h) 20.5 FPS. We also examined the effect of frame count (arrows labeled 'Add frames'). The transition from the single-frame CNN to the transformer processing spatio-temporal data (video) ((b) to (d)) improved suppression of the false molecular signals from the cardiac chambers, although residual signals remained. Increasing the frame count further enhanced suppression of free-floating MB signals, and the increase from 16 to 32 frames additionally produced a slight background denoising effect. Overall, every component combination from SegFormer3D toward MBFormer improved performance over the baseline while maintaining a competitive frame rate, and MBFormer (32 frames) achieved the best balance across all metrics, confirming the effectiveness of each proposed modification.

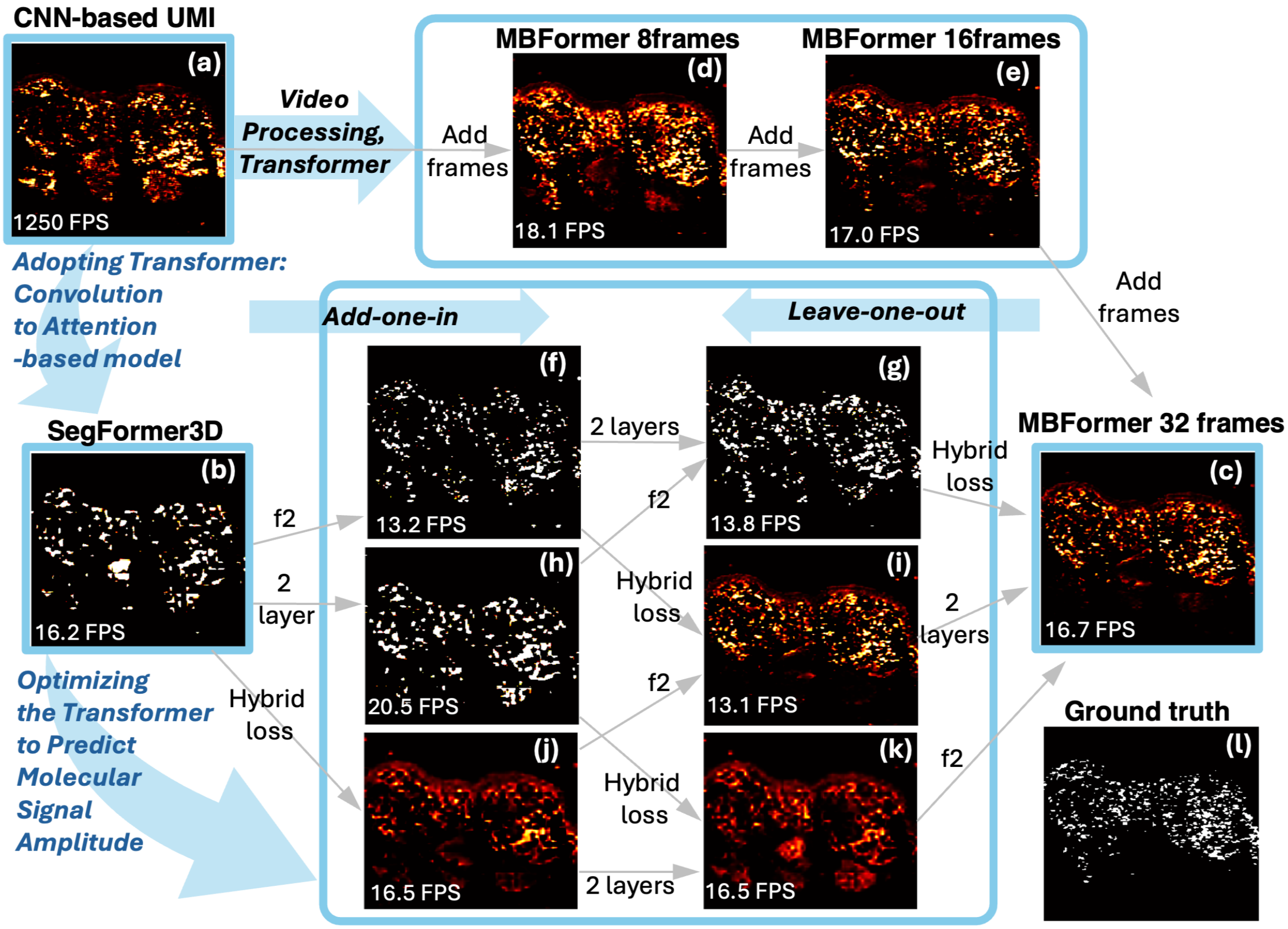

**Figure 5**. Visualization of ablation study. Molecular signal maps illustrate the transition from (a) the baseline CNN-based UMI and (b) SegFormer3D toward (c) MBFormer through Add-one-in and Leave-one-out modifications. Arrow denote individual component updates: ‘f2’, downsampling factor of 4 to 2; ‘Hybrid loss’, loss function update from Dice loss to hybrid loss; ‘2 layers’, hierarchical structure update from 4 layers to 2 layers; and ‘Add frames’, increase in the number of frames utilized for spatio-temporal 3D data (video) processing (8, 16, and 32 frames). Panels (d)–(k) show intermediate configurations, and (l) shows the ground truth. The frame rate (FPS) is displayed at the bottom-left of each panel.

## C. Suppression of free-floating MBs

To verify that MBFormer can effectively suppress free-floating MB signals, we computed CC between consecutive frames (Fig. 3 and Table 2), and furthermore analyzed molecular signal difference between consecutive frames in Fig. 6 (a), molecular signal amplitude profile in Fig. 6 (b) and Fig. 7, and molecular signal amplitude as a function of free-floating MB content in Fig. 8.

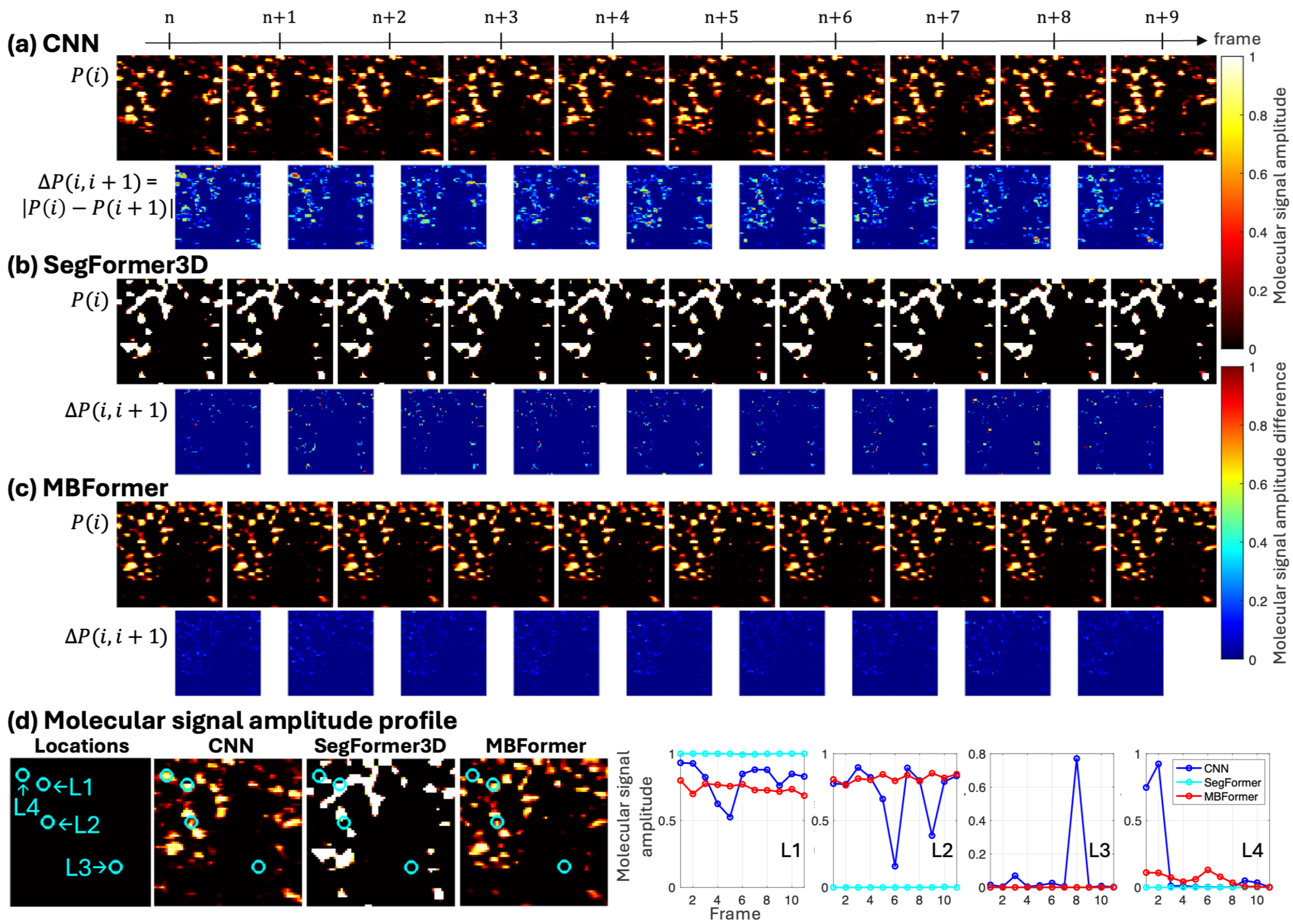

**Figure 6.** Comparison of free-floating (unbound) MB suppression in video processing. Molecular signal amplitude ($P(i)$) at frames n to n+9 with the corresponding inter-frame difference maps ($\Delta P(i, i+1)$) for (a) CNN-based UMI, (b) SegFormer3D, and (c) MBFormer. (d) Molecular signal amplitude profiles over time for four locations: (L1) bound MB with large texture, (L2) bound MB with small texture, and (L3, L4) free-floating MBs passing the location as frames 8 and 2, respectively.

Figs. 6 (a-c) illustrate the molecular signal difference maps between consecutive frames ($\Delta P(i, i+1)$) for CNN, SegFormer3D, and MBFormer, respectively. MBFormer exhibited the lowest inter-frame difference, whereas CNN showed the highest. Fig. 6 (d) compares the molecular signal amplitude over time at four representative locations, denoted by cyan circles labeled L1, L2, L3, and L4 in the left-most image. The signals from MBFormer and SegFormer3D were more stable and consistent across frames than those from CNN. L1 indicates a relatively large MB texture, whereas L2 indicates relatively small. L3 and L4 indicate a location where free MBs passed. The stable profiles of MBFormer at L1 and L2 indicate that MBFormer reliably detects both large (L1) and small (L2) textures of bound MBs, whereas CNN showed time-varying detections due to non-stationary free-floating MB signals. SegFormer was capable of detecting the larger MB texture (L1) but failed to detect the small texture (L2). Based on the CNN profiles at locations L3 and L4, the free-MBs passed there at frames 8 and 2, respectively. CNN detected the signals from free-floating MBs, while MBFormer and SegFormer3D successfully suppressed these signals.

Fig. 7 (a) illustrates the experimental procedure and the binding behavior of targeted MBs. Upon injection, MBs circulated, with targeted MBs adhering to binding sites, while non-adherent free-floating MBs were gradually washed out, ultimately leaving only bound MBs within the lesion. Fig. 7 (b) shows the corresponding temporal profiles, representing the average molecular signal amplitude within the lesion over time, where $t_1$ and $t_2$ represent time (s) after injection and after restart recording after pause for free-MB washing out, respectively. The CNN profile exhibited an initial increase corresponding to the bolus injection, followed by a decrease and then a relatively stable molecular signal after the unbound free-floating MBs washed out. In contrast, MBFormer profile showed a slower increase as the targeted MBs bound to their sites, followed by a saturated profile once all binding sites were occupied. Before the recording pause, the CNN profile showed higher molecular signal compared to MBFormer, whereas after washout, the CNN and MBFormer profiles were nearly equivalent. This suggests that CNN detects both free-floating and bound MBs, whereas MBFormer captures only bound MBs. Fig. 7 (c) shows the molecular images generated by each method for selected time points: $t_1$ = 33, 100, and 157s; $t_2$ = 20s. MBFormer and SegFormer3D produced stable predictions over time, whereas CNN showed more significant changes over time due to the influx and washout of free-floating MBs, except at $t_2$ = 20s, where only bound MBs remain after the free-floating MBs have been washed out. SegFormer3D detected relatively lower molecular signal than MBFormer (Fig. 7 (b)), because it missed detecting small MB textures as shown in Fig. 7 (c).

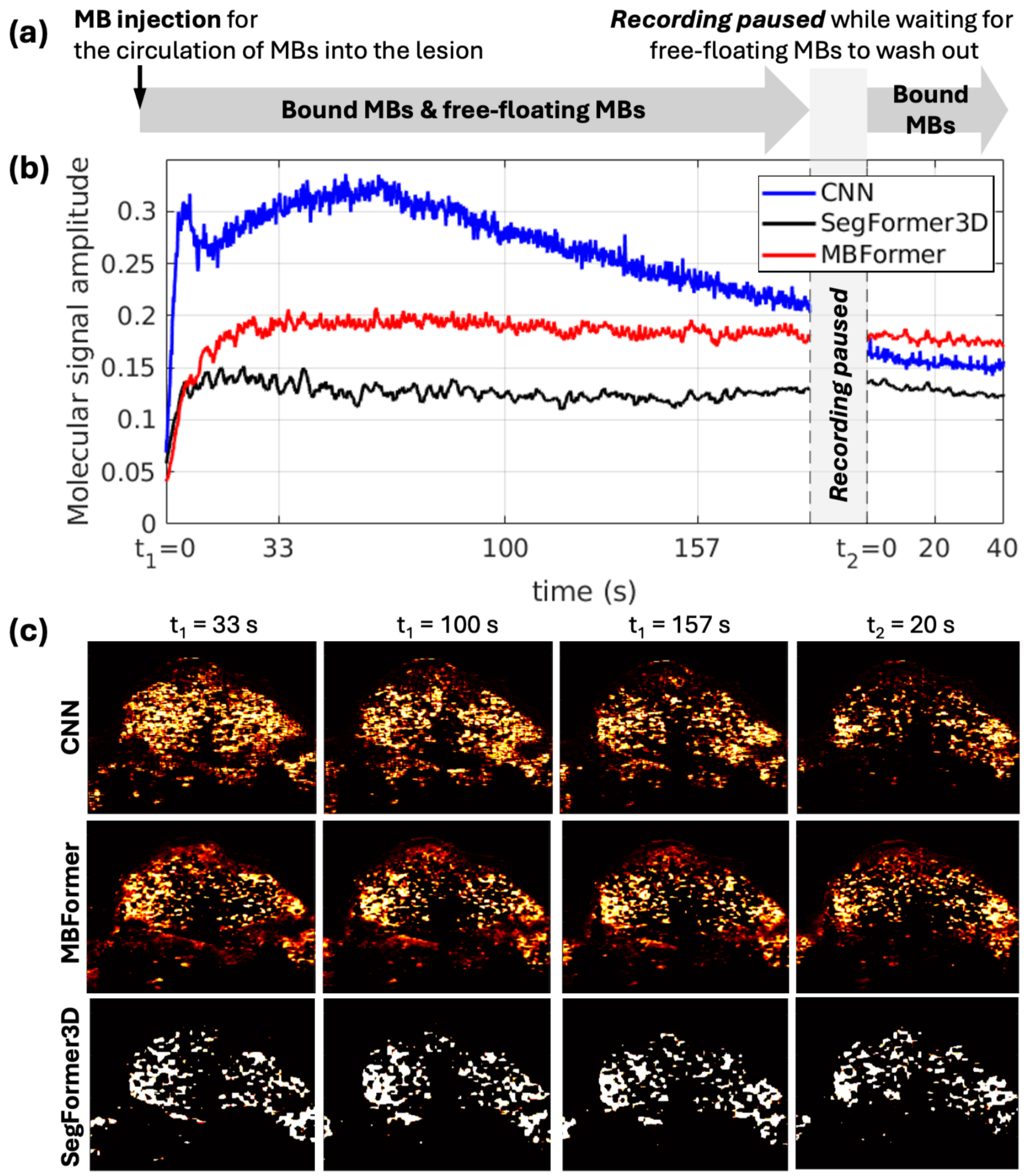

**Figure 7.** Comparison of MBFormer, CNN, and SegFormer3D in detecting bound MBs and free-floating MBs. (a) Overview of our experimental procedures, including MB injection and video recording, alongside the corresponding MB binding behavior. (b) Averaged molecular signal amplitude profile within the lesion for each method. The x-axis indicates time, where $t_1$ and $t_2$ represent the time since MB injection and since recording restart after the washout pause, respectively, aligned with time points in (a). CNN-based UMI detects both free-floating and bound MBs, producing an initial increase after injection due to MB influx, followed by a decrease as unbound MBs wash out, ultimately reaching a stable profile representing only bound MBs. In contrast, MBFormer captures signals exclusively from bound MBs, showing a slower increase as MBs adhere to binding sites and a saturated profile once all binding sites are occupied, without displaying time-varying signals from free-floating MBs. (c) Molecular signal maps at $t_1$ = 33, 100, and 157 s and $t_2$ = 20 s. MBFormer produced almost identical maps over time, whereas CNN detected more MBs in the presence of free-floating MBs.

To further evaluate free-floating MB suppression, we quantified the molecular signal detected by CNN and MBFormer within the lesion and plotted it against the amount of free-floating MBs (Fig. 8). The amount of free-floating MBs was estimated by averaging the intensity of the sparse image obtained through RPCA, and the detected molecular signal was computed by averaging the amplitude within the lesion boundary. Each point represents the average molecular signal across frames for each video, obtained from 99 videos: 17 pre-injection, 39 post-injection containing free-floating MBs, and 43 post-injection videos after free-MB washout. As shown in Fig. 8, when free-floating MBs were scarce, CNN and MBFormer yielded comparable measurements. However, as the amount of free-floating MBs increased, the difference between the two grew: the CNN additionally detected free-floating MBs, whereas MBFormer primarily suppressed these signals.

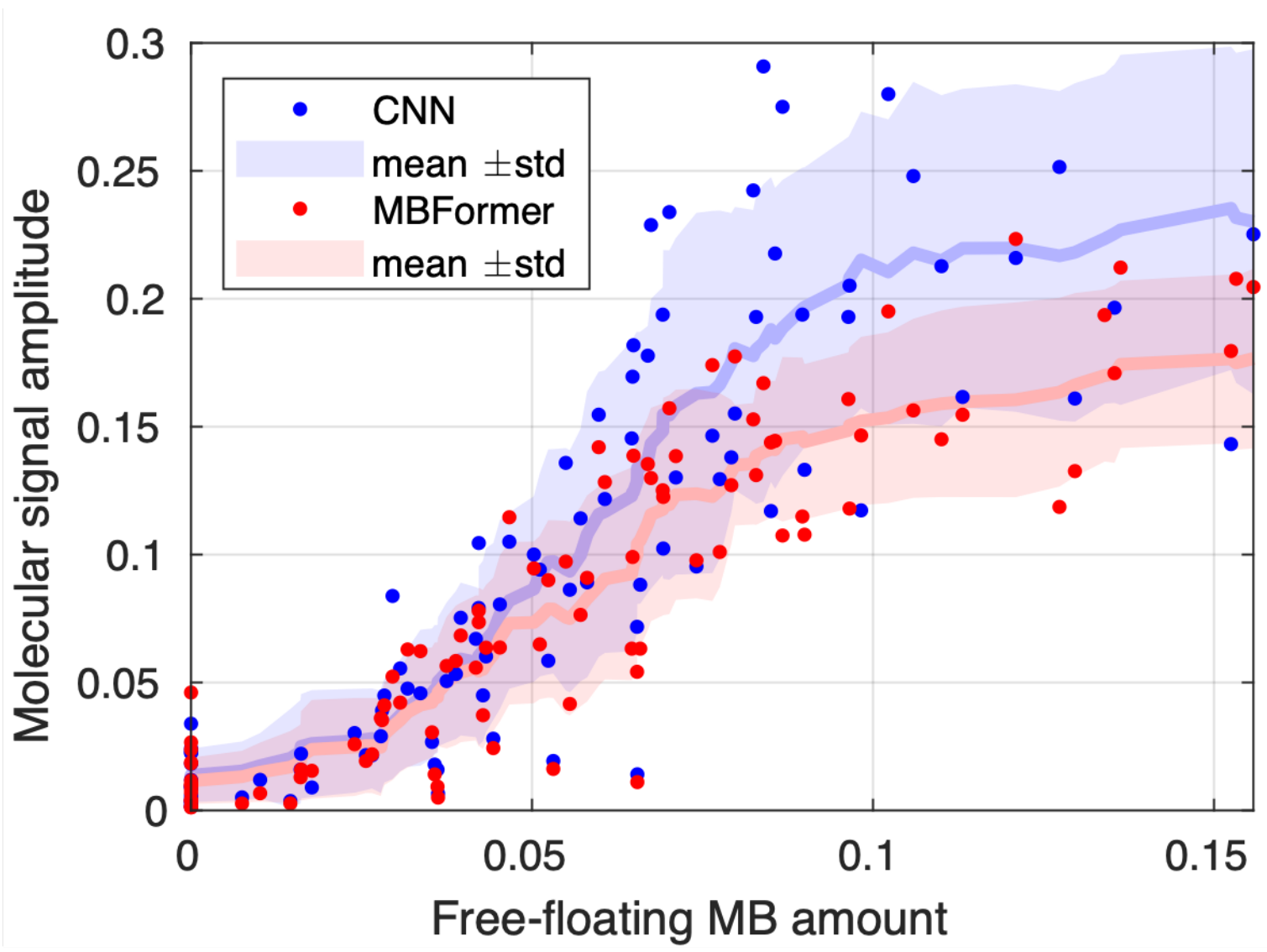

**Figure 8.** Comparison of free-floating MB suppression between CNN-based UMI and MBFormer. The y-axis represents the averaged molecular signal amplitude within the lesion, and the x-axis denotes the amount of free-floating MBs (mean intensity of the sparse RPCA image). Each dot represents one video.

## 6. Discussion

We proposed MBFormer, a transformer-based model for nondestructive UMI that processes 3D spatio-temporal ultrasound video to enhance bound MB detection while suppressing free-floating MBs. The model presents three main contributions. First, by leveraging the distinct temporal behaviors of bound and free-floating MBs, MBFormer selectively preserves stationary bound MB signals and removes free-floating MBs, allowing molecular signals to be attributed more specifically to targeted cancer biomarkers. Second, its attention-based architecture improves prediction of molecular signals from tiny and fine textures characteristic of bound MBs. Third, rather than defining UMI as binary MB segmentation, MBFormer directly estimates the continuous molecular signal amplitude, better capturing the fine intensity variations of tiny MBs. These contributions are realized on a SegFormer3D backbone [18], which we adapted for spatio-temporal video processing through three architectural modifications: a downsampling factor of 2, a reduced hierarchical structure of two layers, and a hybrid loss. MBFormer achieved the highest PCC and AUC, compared with the SegFormer3D baseline and our prior CNN-based UMI [13, 14]; the CNN demonstrated superior performance to the current state-of-the-art UMI modality, namely DTE. Moreover, MBFormer achieved 16.7 – 18.1 FPS, thereby demonstrating a favorable balance between accuracy and real-time capability.

A key advantage of MBFormer, compared to our previous work [13, 14], is its ability to suppress signals from free-floating MBs. The single-frame inference approach of the CNN-based models was limited in its ability to suppress unbound MBs, as it could not assess the temporal information that reflects the time-varying nature of signals generated by free-floating MBs. By processing consecutive frames from video input, MBFormer exploits the temporal behavior of MBs: bound MBs remain stationary across frames, whereas free-floating MBs generate non-stationary time-varying signals and spatial movement between frames. This is reflected in the high inter-frame consistency (CC $\geq$ 0.92) achieved by all video processing models, in contrast to the single-frame CNN (CC = 0.66). Our temporal analysis (Figs. 6-8) further showed that CNN detected both free-floating and bound MBs, producing time-varying signals across consecutive frames due to MB movement (Fig. 6 (d)) and time-varying signal profile during MB influx and washout (Fig. 7 (b)), whereas MBFormer captured only bound MBs, yielding stable predictions over time. Moreover, as free-MB content increased, molecular signal difference between MBFormer and CNN increased (Fig. 8). This selective detection of bound MBs is clinically important because it enables more specific cancer biomarker detection through noninvasive UMI.

Moreover, adopting attention mechanism enabled more accurate detection of molecular signals than the CNN-based UMI approach, as shown in Fig. 2; MBFormer's predictions more closely resembled the DTE and ground truth, particularly for tiny MBs. The CNN-based network extracts features through convolution, which emphasizes local information within convolutional kernels. Although the effective receptive field grows as features pass through consecutive convolutional layers, long-range spatial dependencies are captured only indirectly and progressively. In contrast, the attention-based network models global relationships directly by computing self-attention among overlapping spatio-temporal patch tokens across the entire video input. This allows MBFormer to integrate information over the full spatio-temporal context in a single operation at each hierarchical layer, which may contribute to preserving the fine texture of tiny MBs. Overall, MBFormer produced molecular signal predictions more closely aligned with the DTE than the CNN.

Beyond the architectural design, the choice of prediction head also plays a critical role in UMI. Typical medical imaging segmentation tasks employ a segmentation head to produce binarized outputs. However, binarized prediction easily overlooks relatively weak MB signals that originate from lower concentrations of MBs as well as the “darker speckles” resulting from destructive interference from the scattering of higher concentrations of sub-wavelength MBs. Thus, to detect fine MB textures, MBFormer’s prediction head directly outputs molecular signal amplitude. In addition to the replacement of the prediction head, in our ablation study, the SegFormer3D backbone [18] was modified to process ultrasound videos; the SegFormer3D was originally designed for 3D volumetric medical image segmentation, validated using Brain Tumor Segmentation (BraTS) for magnetic resonance imaging [35] and Synapse Multi-Organ Segmentation (Synapse) for computed tomography imaging [36]. Therefore, the baseline SegFormer3D in our ablation study (Table 2) does not correspond to the original SegFormer3D model, which incorporates both prediction head and video processing. Table 2 demonstrates that the SegFormer3D model exhibits poorer performance in MB detection compared to the results originally reported in [18] for volumetric segmentation of BraTS and Synapse datasets. However, this diminished performance is specific to MB detection and does not indicate that the network inherently underperforms. This gap reflects the domain shift from volumetric organ segmentation to fine-grained MB amplitude estimation considering temporal dynamics of MBs, rather than a limitation of the architecture itself. This motivates our task-specific modifications of the architecture. In this work, we designed a transformer-based architecture for the detection of the fine texture of molecular imaging, while leveraging the advantages of SegFormer, including its lightweight MLP decoder and positional embedding-free design [29], which contributes to faster processing.

The choice of loss function critically affected both molecular signal prediction and its evaluation. In our ablation study, models trained with the Dice loss (CDC = 0.82 – 0.83) resulted in higher CDC than those trained with the hybrid loss, as provided in Table 2. However, this higher CDC did not always reflect superior performance. As shown in Fig. 5, the Dice loss models (Figs. 5 (b, f-h)) yielded near-binary maps, whereas the hybrid loss models produced continuous outputs that better preserved the fine MB textures. Because CDC is a region-overlap metric computed against a binary ground truth, near-binary predictions inflate the CDC while failing to capture the underlying molecular signal, as reflected by their lower PCC than MBFormer models. In contrast, the hybrid loss, combining the Dice with a cross-entropy loss, discouraged overly binarized outputs and preserved continuous values that better captured fine molecular signal textures, yielding the relatively higher PCC. This indicates that CDC alone is insufficient metric for our task, and that PCC, which directly compares amplitude values, is a more appropriate metric.

Given the importance of these continuous outputs, we refer to the predictions as the molecular signal amplitude rather than a probability. Although the outputs are produced by a softmax function and range between 0 and 1, an uncalibrated softmax output does not represent a true probability without calibration such as temperature scaling [37]. When we applied temperature scaling, calibration shifted and amplified low-valued predictions, producing noise-like background artifacts in the predicted maps. Such artifacts could be removed by thresholding after calibration, but this would make the resulting images threshold-dependent. More importantly, our UMI goal is not to estimate calibrated probability but to visualize bound MBs for lesion detection. Therefore, we report the softmax output directly as the molecular signal amplitude, preserving diagnostically relevant bound MB signals.

Our measurement of the inference time for MBFormer, using two GPUs in parallel, yielded rates between 16.7 and 18.1 FPS, demonstrating the feasibility of real-time implementation. Given that our current implementation is a preliminary model, we anticipate that further code optimization could enhance both the FPS and overall model efficiency. Additionally, enhancing the MBFormer architecture or devising new networks specifically tailored for faster inference could facilitate the transition of nondestructive UMI to real-time, free-hand applications within clinical environments.

We validated our prior CNN-based UMI [13] and deployed it on a research ultrasound system (Vantage 256 ultrasound system, Verasonics Inc, Kirkland, WA) [12]. The real-time implementation of CNN-based UMI, which showed an inference frame rate of 1250 FPS, achieved a frame rate of 20-30 FPS when including the round-trip pulse-echo time, MATLAB-based ultrasound beamforming, signal processing, and image reconstruction of B-mode, CEUS, and UMI images, as well as the integration of the TensorFlow CNN model into the Verasonics MATLAB environment. This significant decrease in frame rate when deployed on an ultrasound system suggests that the current inference time of MBFormer (between 16.7 and 18.1 FPS) may pose a challenge for achieving real-time performance. Therefore, while this work successfully achieved bound MB detection/free-bubble suppression and demonstrated the potential of employing attention-based UMI with feasible inference times for clinical translation, it remains essential to explore faster network architectures that preserve the detection of bound MB and suppression of free-floating MBs that was demonstrated in this study. It is important to note that the significant increase in runtime with the CNN model is also attributed to the hardware environment of the research ultrasound system; thus, other commercial machines most likely will achieve faster runtimes due to optimized data transfer pipelines, dedicated image reconstruction hardware, and software implementations that could integrate such neural network models.

## 7. Conclusion

We developed a transformer-based nondestructive UMI, named MBFormer, to achieve enhanced suppression of free-floating MBs and more accurate detection of the signals from bound MBs. MBFormer was trained and validated using an *in vivo* study of a transgenic mouse model of breast cancer, in which its performance in detecting bound MBs, suppressing unbound MBs, and runtime was compared against a prior CNN model and the SegFormer3D baseline. The use of 3D spatio-temporal (ultrasound videos) processing markedly improved the suppression of unbound MBs compared with the single-frame processing of the CNN-based model. Moreover, the attention-based transformer enhanced the visualization of fine MB textures, and the hybrid loss enabled a continuous representation of the molecular signal amplitude. By developing MBFormer upon a lightweight SegFormer3D backbone and tailoring the architecture for fine MB texture detection, we achieved an inference time of 55.2 to 59.8 ms, corresponding to a frame rate of 16.7 to 18.1 FPS, demonstrating the potential for real-time implementation while preserving stable detection of fine MB textures. Therefore, we anticipate that this improved nondestructive UMI framework, leveraging attention mechanisms, can potentially be integrated into clinical systems for accurate detection of bound MBs, suppression of unbound MBs, and real-time and free-hand imaging.

*Acknowledgments*— This work was supported by grant R01-EB031799 from the National Institute of Biomedical Imaging and Bioengineering and grant R01-CA218204 from the National Cancer Institute.

## References

[1] J. Bzyl, W. Lederle, A. Rix *et al.*, "Molecular and functional ultrasound imaging in differently aggressive breast cancer xenografts using two novel ultrasound contrast agents (BR55 and BR38)," *European radiology,* vol. 21, pp. 1988-1995, 2011.

[2] S. V. Bachawal, K. C. Jensen, A. M. Lutz *et al.*, "Earlier detection of breast cancer with ultrasound molecular imaging in a transgenic mouse model," *Cancer research,* vol. 73, no. 6, pp. 1689-1698, 2013.

[3] S. V. Bachawal, K. C. Jensen, K. E. Wilson *et al.*, "Breast cancer detection by B7-H3–targeted ultrasound molecular imaging," *Cancer research,* vol. 75, no. 12, pp. 2501-2509, 2015.

[4] J. K. Willmann, L. Bonomo, A. C. Testa *et al.*, "Ultrasound molecular imaging with BR55 in patients with breast and ovarian lesions: first-in-human results," *Journal of Clinical Oncology,* vol. 35, no. 19, pp. 2133, 2017.

[5] D. H. Simpson, C. T. Chin, and P. N. Burns, "Pulse inversion Doppler: a new method for detecting nonlinear echoes from microbubble contrast agents," *IEEE transactions on ultrasonics, ferroelectrics, and frequency control,* vol. 46, no. 2, pp. 372-382, 1999.

[6] P. Phillips, "Contrast pulse sequences (CPS): imaging nonlinear microbubbles." pp. 1739-1745.

[7] R. J. Eckersley, C. T. Chin, and P. N. Burns, "Optimising phase and amplitude modulation schemes for imaging microbubble contrast agents at low acoustic power," *Ultrasound in medicine & biology,* vol. 31, no. 2, pp. 213-219, 2005.

[8] J. R. Lindner, J. Song, J. Christiansen *et al.*, "Ultrasound assessment of inflammation and renal tissue injury with microbubbles targeted to P-selectin," *Circulation,* vol. 104, no. 17, pp. 2107-2112, 2001.

[9] P. A. Dayton, and J. J. Rychak, "Molecular ultrasound imaging using microbubble contrast agents," *Front Biosci,* vol. 12, no. 23, pp. 5124-5142, 2007.

[10] J. r. K. Willmann, R. Paulmurugan, K. Chen *et al.*, "US imaging of tumor angiogenesis with microbubbles targeted to vascular endothelial growth factor receptor type 2 in mice," *Radiology,* vol. 246, no. 2, pp. 508-518, 2008.

[11] M. Dl, "Bioeffects considerations for diagnostic ultrasound contrast agents," *J Ultrasound Med,* vol. 27, pp. 611-632, 2008.

[12] H. S. Hashemi, D. Hyun, N. Nguyen *et al.*, "Enhancing ultrasound molecular imaging: Toward real-time RPCA-based filtering to differentiate bound and free microbubbles," *IEEE transactions on ultrasonics*, 2025.

[13] J. Baek, D. Hyun, A. Natarajan *et al.*, "Improved Nondestructive Ultrasound Molecular Imaging with Lightweight Convolutional Neural Network," *IEEE Transactions on Medical Imaging*, 2026.

[14] D. Hyun, L. Abou-Elkacem, R. Bam *et al.*, "Nondestructive detection of targeted microbubbles using dual-mode data and deep learning for real-time ultrasound molecular imaging," *IEEE transactions on medical imaging,* vol. 39, no. 10, pp. 3079-3088, 2020.

[15] E. B. Herbst, S. Unnikrishnan, A. L. Klibanov *et al.*, "Validation of normalized singular spectrum area as a classifier for molecularly targeted microbubble adherence," *Ultrasound in medicine & biology,* vol. 45, no. 9, pp. 2493-2501, 2019.

[16] G. Collado-Lara, G. Wahyulaksana, H. J. Vos *et al.*, "Nondestructive Ultrasound Molecular Imaging With Higher Order Singular Value Decomposition," *IEEE*

*Transactions on Ultrasonics, Ferroelectrics, and Frequency Control,* vol. 72, no. 8, pp. 1095-1107, 2025.

[17] J. Baek, D. Hyun, A. Natarajan *et al.*, "Nondestructive ultrasound molecular imaging based on a neural network approach utilizing post-processed ultrasound images." pp. 1-3.

[18] S. Perera, P. Navard, and A. Yilmaz, "Segformer3d: an efficient transformer for 3d medical image segmentation." pp. 4981-4988.

[19] O. Ronneberger, P. Fischer, and T. Brox, "U-net: Convolutional networks for biomedical image segmentation." pp. 234-241.

[20] M. Byra, P. Jarosik, A. Szubert *et al.*, "Breast mass segmentation in ultrasound with selective kernel U-Net convolutional neural network," *Biomedical signal processing and control,* vol. 61, pp. 102027, 2020.

[21] R. Almajalid, J. Shan, Y. Du *et al.*, "Development of a deep-learning-based method for breast ultrasound image segmentation." pp. 1103-1108.

[22] A. Vaswani, N. Shazeer, N. Parmar *et al.*, "Attention is all you need," *Advances in neural information processing systems,* vol. 30, 2017.

[23] A. Dosovitskiy, L. Beyer, A. Kolesnikov *et al.*, "An image is worth 16x16 words: Transformers for image recognition at scale," *arXiv preprint arXiv:2010.11929*, 2020.

[24] A. Arnab, M. Dehghani, G. Heigold *et al.*, "Vivit: A video vision transformer." pp. 6836-6846.

[25] J. Chen, Y. Lu, Q. Yu *et al.*, "Transunet: Transformers make strong encoders for medical image segmentation," *arXiv preprint arXiv:2102.04306*, 2021.

[26] Z. Liu, Y. Lin, Y. Cao *et al.*, "Swin transformer: Hierarchical vision transformer using shifted windows." pp. 10012-10022.

[27] H. Cao, Y. Wang, J. Chen *et al.*, "Swin-unet: Unet-like pure transformer for medical image segmentation." pp. 205-218.

[28] H.-Y. Zhou, J. Guo, Y. Zhang *et al.*, "nnformer: Interleaved transformer for volumetric segmentation," *arXiv preprint arXiv:2109.03201*, 2021.

[29] E. Xie, W. Wang, Z. Yu *et al.*, "SegFormer: Simple and efficient design for semantic segmentation with transformers," *Advances in neural information processing systems,* vol. 34, pp. 12077-12090, 2021.

[30] A. Hatamizadeh, Y. Tang, V. Nath *et al.*, "Unetr: Transformers for 3d medical image segmentation." pp. 1748-1758.

[31] Y. Xie, J. Zhang, C. Shen *et al.*, "Cotr: Efficiently bridging cnn and transformer for 3d medical image segmentation." pp. 171-180.

[32] S. Zheng, J. Lu, H. Zhao *et al.*, "Rethinking semantic segmentation from a sequence-to-sequence perspective with transformers." pp. 6877-6886.

[33] W. Wang, E. Xie, X. Li *et al.*, "Pyramid vision transformer: A versatile backbone for dense prediction without convolutions." pp. 568-578.

[34] R. R. Shamir, Y. Duchin, J. Kim *et al.*, "Continuous dice coefficient: a method for evaluating probabilistic segmentations," *arXiv preprint arXiv:1906.11031*, 2019.

[35] B. H. Menze, A. Jakab, S. Bauer *et al.*, "The multimodal brain tumor image segmentation benchmark (BRATS)," *IEEE transactions on medical imaging,* vol. 34, no. 10, pp. 1993-2024, 2014.

[36] B. Landman, Z. Xu, J. Igelsias *et al.*, "Miccai multi-atlas labeling beyond the cranial vault–workshop and challenge." p. 12.

[37] C. Guo, G. Pleiss, Y. Sun *et al.*, "On calibration of modern neural networks." pp. 1321-1330.